\documentclass[
reprint,
superscriptaddress,
footinbib,
amsmath,amssymb,
aps,
prl
]{revtex4-1}
\usepackage{graphicx}
\usepackage{array}
\usepackage{dcolumn}
\usepackage{bm}
\usepackage[utf8]{inputenc}
\usepackage[english]{babel}
\usepackage{amsmath}    
\usepackage{graphicx}   
\usepackage{verbatim}   
\usepackage[usenames, dvipsnames]{color}      
\usepackage{subfigure}  
\usepackage{hyperref}   
\hypersetup{colorlinks=true,citecolor=blue,linkcolor=red,urlcolor=blue}
\usepackage{color}

\definecolor{green}{rgb}{0,0.5,0}

\newcommand{\B}[1]{{\bm{#1}}}

\begin{document}
\title{Elementary plastic events in amorphous silica}
\author{Silvia Bonfanti$^1$, Roberto Guerra$^1$, Chandana Mondal$^2$, Itamar Procaccia$^{2,3}$ and Stefano Zapperi$^{1,4}$ }
\affiliation{ $^1$Center for Complexity and Biosystems, Department of Physics, University of Milan, via Celoria 16, 20133 Milano, Italy\\  $^2$ Dept. of Chemical Physics, The Weizmann Institute of Science, Rehovot 76100, Israel\\ $^3$Center for OPTical IMagery Analysis and Learning, Northwestern Polytechnical University, Xi'an, 710072 China\\ $^4$ CNR - Consiglio Nazionale delle Ricerche,  Istituto di Chimica della Materia Condensata e di Tecnologie per l'Energia, Via R. Cozzi 53, 20125 Milano, Italy}
\date{\today}

\date{\today}

\begin{abstract}
Plastic instabilities in amorphous materials are often studied using idealized
models of binary mixtures that do not capture accurately molecular interactions
and bonding present in real glasses. Here we study atomic scale plastic instabilities in
a three dimensional molecular dynamics model of silica glass under quasi-static shear.
We identify two distinct types of elementary plastic events, one is a standard quasi-localized atomic
rearrangement while the second is a bond breaking event that is absent in
simplified models of fragile glass formers. Our results show that both plastic events can be predicted
by a drop of the lowest non-zero eigenvalue of the Hessian matrix that vanishes at a critical strain.
Remarkably, we find  very high correlation between the associated eigenvectors and the non-affine displacement
fields accompanying the bond breaking event, predicting the locus of structural failure. Both eigenvectors and
non-affine displacement fields display an Eshelby-like quadrupolar structure
for both failure modes, rearrangement or bond-breaking. Our results thus clarify the
nature of atomic scale plastic instabilities in silica glasses providing useful information
for the development of mesoscale models of amorphous plasticity.
\end{abstract}
\maketitle

\begin{figure*}
\includegraphics[width=0.40\textwidth]{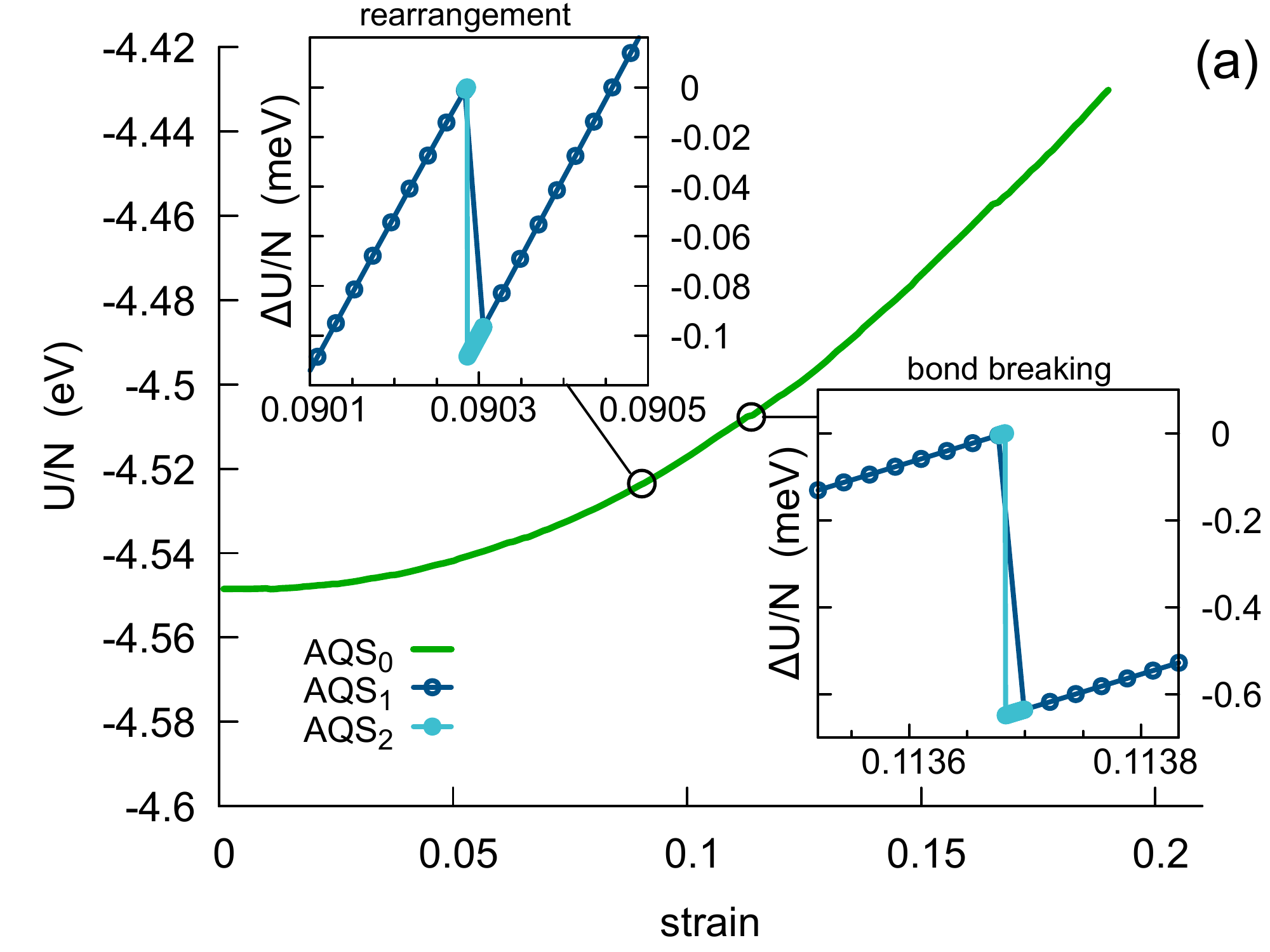}~~~~~~~
\includegraphics[width=0.50\textwidth]{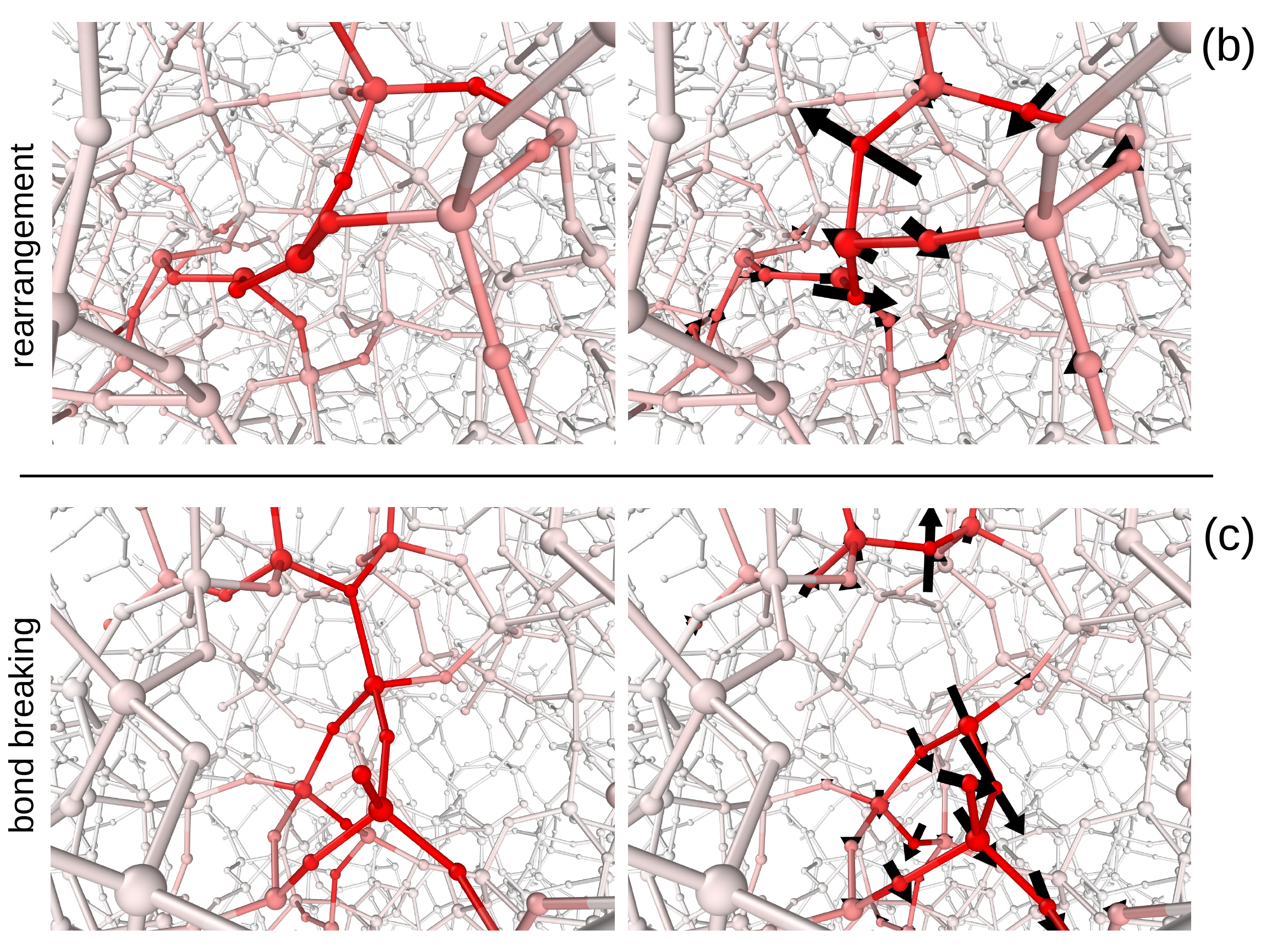}
\caption{(a) Typical dependence of potential energy per particle $U/N$ on strain in the AQS$_0$ protocol for the 3D silica glass. The insets are a blowup of two events emphasized by the circles on the main curve, the first being a localized rearrangement (top) and the second (bottom) reporting the first observed bond breaking (Si--O bond). Blue and light blue curves are obtained by decreasing strain step size, AQS$_1$ and AQS$_2$ respectively. The energy zero has been set to the maximum value before the drop, which is larger in the bond breaking event. (b),(c) 3D perspective view of the atomic positions before (left) and after (right) the energy drop for the corresponding events of panel (a). Large and small spheres represent Si and O atoms, respectively. Color and arrows report the magnitude of the occurred non-affine atomic displacements. Arrows have been rescaled by a factor of $2$. The corresponding movies are available in Supplemental Material \cite{Note21}.}
\label{fig:displacement_snap}
\end{figure*}

Amorphous solids under applied shear deformation undergo localized plastic instabilities associated with the rearrangement of a subset of particles and an associated
energy release. These particle reorganization induces structural deformation patterns,
that have been identified experimentally and numerically in amorphous materials such as silica glasses~\cite{horbach1996finite,coslovich2009dynamics,huang2013imaging}, metallic glasses~\cite{argon1979plastic}, colloidal glasses~\cite{chikkadi2011long}, foams~\cite{dennin2004statistics}, bubble rafts~\cite{argon1979plastic}, and emulsions~\cite{hebraud1997yielding,clara2015affine}.
The initial destabilization can give rise to a progression of
additional deformation events in some other areas of the sample, up to the global material failure. The ability to predict the plastic instabilities and characterize their spatial features is of fundamental importance to understand the mechanical response of amorphous solids
and to devise mesoscale model focusing on the evolution of localized plastic events \cite{baret2002extremal,budrikis2013avalanche,budrikis2017universal,nicolas2018deformation}.

A useful theoretical framework to analyze elementary plastic events is the limit of temperature $T=0$ and of quasistatic strain where the real space structure can be easily related with a potential energy landscape description~\cite{stillinger1995topographic}. To this end, many computational studies on amorphous solids have been performed with athermal quasi static (AQS) protocol~\cite{maloney2006amorphous,malandro1999relationships,11HKLP,12DKP,12DHP}:  a glass sample initially quenched down to zero temperature is deformed by a quasi static shear procedure
consisting in the relaxation of the system after each strain step.
Within the AQS conditions the elastic and plastic features of amorphous solids can be understood by analyzing the Hessian matrix
\begin{equation}
H_{ij} \equiv \frac{\partial^2 U(\B r_1, \cdots ,\B r_N)}{\partial \B r_i \partial \B r_j} \equiv - \frac{\partial \B f_{i}(\B r_1, \cdots ,\B r_N) }{\partial \B r_j}
\label{defHes}
\end{equation}
where $U(\B r_1,\B r_2, \cdots \B r_N)$ is the total potential energy of the system, $\B f_i$ is the force vector on particle~$i$, and $\{ \B r_i\}_{i=1}^N$ are the coordinates of the particles. The explicit $\B H$ element expression is reported in \setcounter{footnote}{20}\footnote{See Supplemental Material at http://link.aps.org/supplemental/xxx for information on the Hessian matrix and movies related to Fig.~1.}.
When the system is mechanically stable the eigenvalues~$\lambda$ of the Hessian are semi-positive, with zero values for the Goldstone modes and all the rest positive.
Elementary plastic instabilities are signaled by the lowest eigenvalue  $\lambda_{min}$ going to zero and an eigenfunction getting quasi-localized on a pattern correlated with real space non-affine displacements. Typically observed quadrupole-like structure can be described as an ellipsoidal inclusion in an elastic medium \cite{PhysRevE.88.032401}, following the classic work of Eshelby \cite{Eshelby1957}. This kind of analysis always gives rise to a similar phenomenology, independently of the detailed microscopic interactions between the constituents \cite{12DKP}, but to the best of our knowledge it has only be applied to idealized model of fragile glasses~\cite{angell1988perspective} such as metallic glasses~\cite{falk1998dynamics,tanguy2006plastic,12DKP} or frictional disks whose packing structure is isotropic~\cite{PhysRevLett.107.108302,SciPostPhys.1.2.016}.

Experimental evidences of atomic rearrangements for two-dimensional silica glass have been reported in Ref.~\cite{huang2013imaging}, while numerically the plastic rearrangements in strained silica at zero temperature has been investigated in Ref. \cite{koziatek2015short}.
We are lacking, however, numerical studies of normal modes in realistic strong glass formers such as silica that is characterized by a strong chemical structural order with tetrahedral networks made by covalent bonds. Indeed silica glass is not only appealing for technological and commercial applications but also for its intriguing and anomalous behavior that is still not fully understood. In particular, we mention here the peaks in the specific heat, the diffusion constant, the density maximum~\cite{guo2018fluctuations} and so on, that differentiates silica from all other fragile glasses. Therefore \textit{a priori} the nature of plastic instabilities in silica glasses is not clear, specially considering the relevance of \textit{anisotropic} bonds which are absent in other well studied amorphous systems.

In this paper, we study three dimensional silica glass under AQS shear conditions. Previous numerical work \cite{doi:10.1021/acs.nanolett.8b00469} has shown that bond breaking is mainly responsible for damage accumulation and failure of silica at zero temperature. In this paper we focus on the initial single events acting as fracture precursors and analyze the softest modes.


We perform simulations on a silica glass sample in a cubic box. The system is formed by a total of $N=8250$ atoms, composed by $N_{Si}=2750$ silicon atoms and $N_O=5500$ oxygen atoms.
Silica glasses are simulated using the Watanabe's potential~\cite{watanabe2004improved} with a similar sample preparation strategy. The advantage of this potential is that the usual Coulomb interaction term is implicitly replaced by a
coordination-based bond softening function for Si-O atoms that accounts for the
environmental dependence, therefore we perform simulations in open boundary conditions to
study surface effects. The general form of the potential consists of two terms: a two-body
interaction that depends on distance and a three-body interaction that describe rotational and
translational symmetry.
The Hessian matrix (Eq.~\ref{defHes}) is computed numerically from the first-order derivatives of inter-particle forces. To this extent, each element $H_{ij}$ is obtained by calculating the force acting on particle $i$ following a displacement of particle $j$ by a small amount, $\delta=10^{-7}$\,\AA~along positive and negative direction, and by applying the difference quotient.
All the simulations have been performed using the {\small LAMMPS} simulator
package \cite{lammps}, and visualized with the {\small OVITO} package \cite{ovito}.

To generate the sample we have started from a randomly positioned Si,O atoms, with density ${\rho}_{in}=2.196$\,g/cm$^3$ in a box size of 5$\times$5$\times$5\,nm$^3$.
We then have applied the following annealing procedure: i) After an initial 2\,ps of NVE dynamics with LJ interaction limited to 1\,\AA/ps, we switch to our reference Watanabe's potential for silica~\cite{watanabe2004improved} and ii) we perform subsequent 8\,ps of NVE dynamics. iii) We then heat up the system up to 6000\,K in 30\,ps, iv) thermalize at 6000\,K for 80\,ps, v) reduce the temperature to 4000\,K in 30\,ps, vi) then to 0.01\,K in 50\,ps, vii) then to 0.001\,K in 100\,ps, and viii) finally we perform a pressure minimization -- cell relaxation -- for 50\,ps. After such procedure we get a final density ${\rho}_{fin}=2.2439$\,g/cm$^3$ and box size 4.948$\times$4.996$\times$4.948\,nm$^3$. Analysis on such initial sample compares well with experimentally observed density~\cite{brueckner1970properties} and with previous calculations of atomic coordination \cite{vollmayr2013temperature}.

The so-produced configuration is then used to start the AQS protocol.
At each AQS step \cite{maloney2006amorphous}, we strain the sample along $z$ and compress it along $x$ and $y$ according to a Poisson ratio $\nu$=0.17.
We have adopted three different increment of strain $\delta\gamma$, namely $\delta\gamma$=5$\cdot$10$^{-4}$ (AQS$_0$), $\delta\gamma$=(5$\cdot$10$^{-4}$)/50 (AQS$_1$), $\delta\gamma$=(5$\cdot$10$^{-4}$)/50/50 (AQS$_2$). In order to reduce the computational burden, once a rearrangement event is identified in the faster AQS$_0$ simulation, we used the more refined AQS$_1$ and then AQS$_2$ simulations only in the vicinity of such event.
After the imposed $\delta\gamma$ strain an energy minimization through the FIRE scheme \cite{bitzek2006structural} is performed until a maximum force of $10^{-10}$\,eV/{\AA} is reached.


\begin{figure}[t]
\centering
\includegraphics[width=0.49\columnwidth]{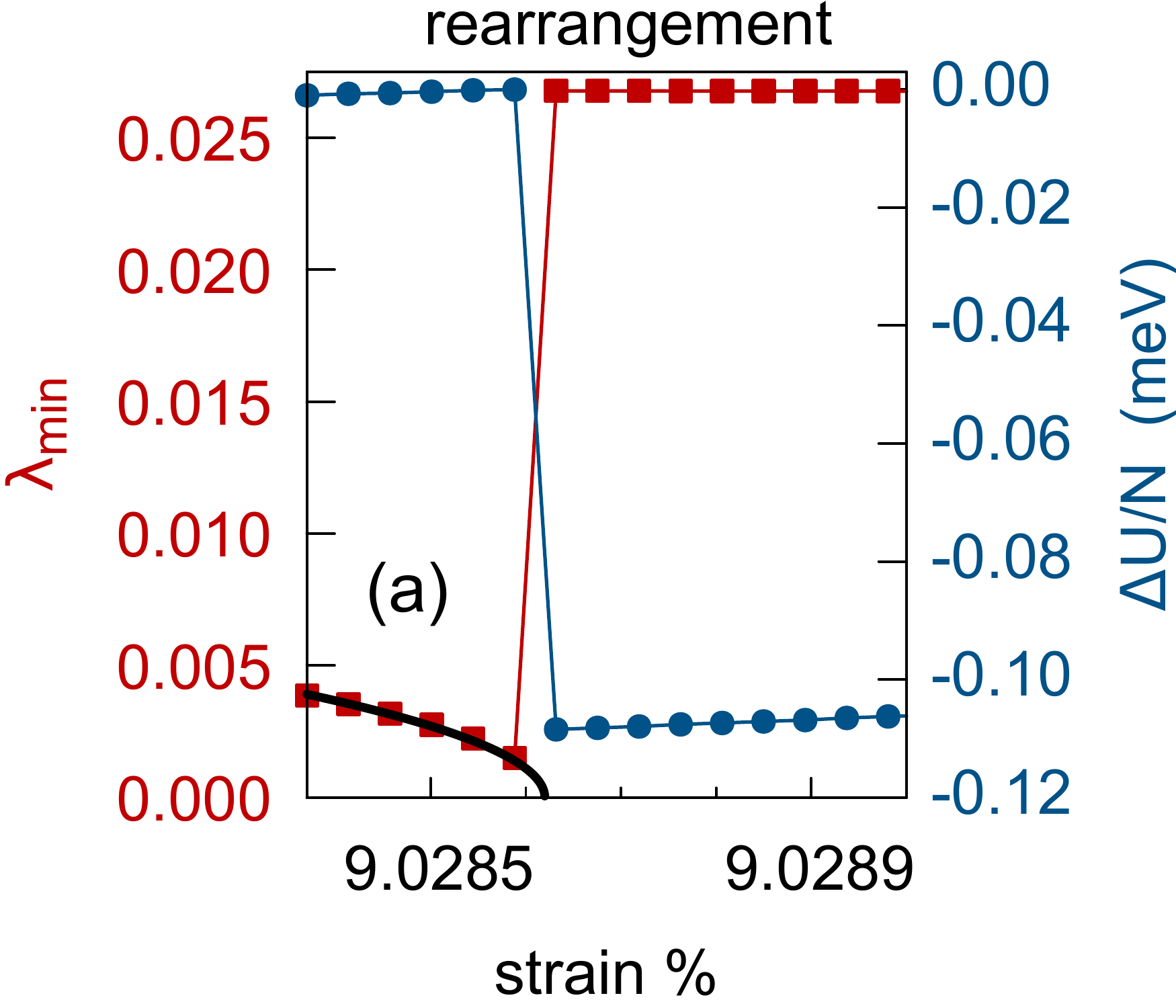}
\includegraphics[width=0.49\columnwidth]{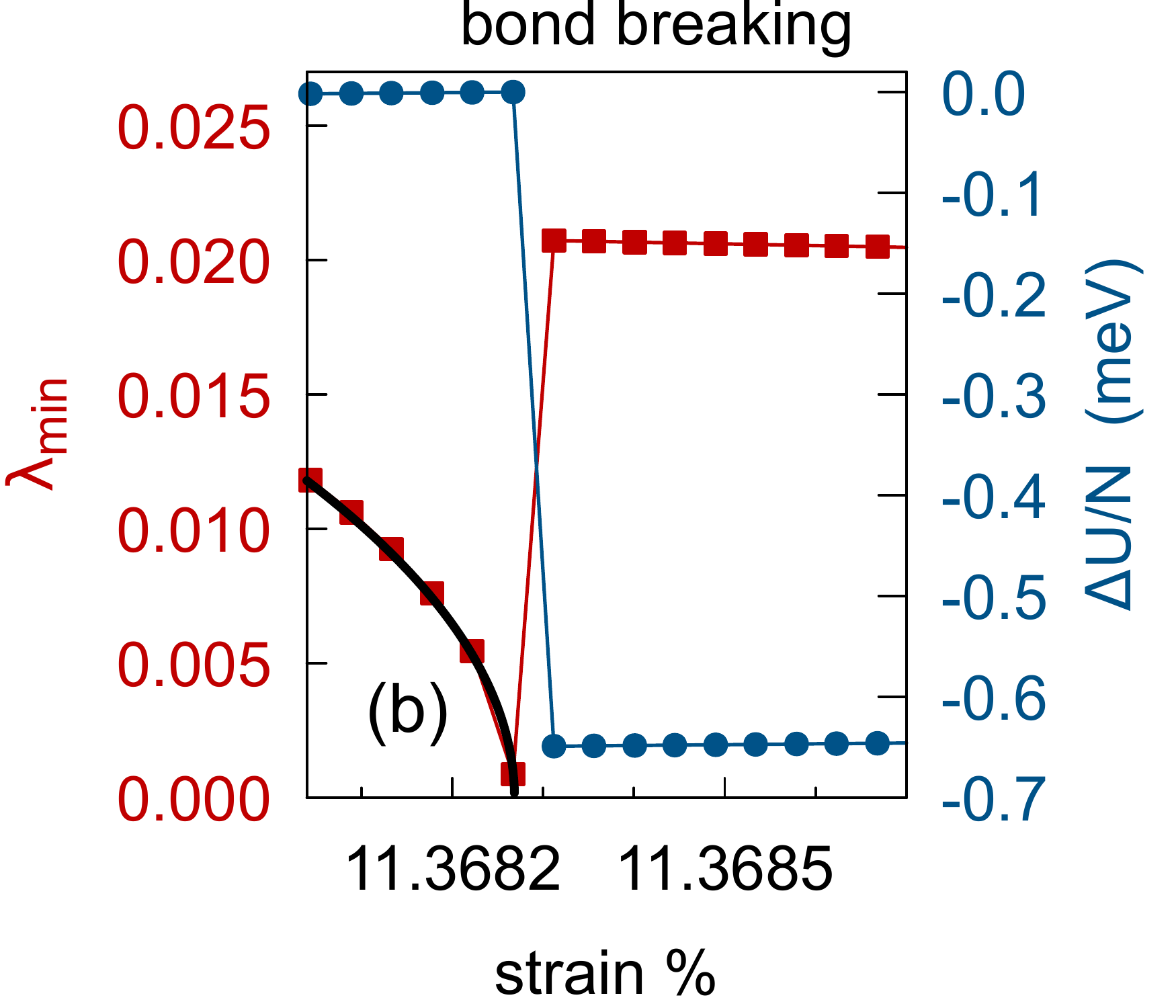}
\caption{Lowest eigenvalues and potential energy trend comparison for AQS$_2$ strain steps for (a) localized atomic rearrangement and (b) bond breaking events. The lowest eigenvalue is reported in red squares and approaches zero at the critical strain $\gamma_c$~$\simeq$\,9.02862\% and $\simeq$11.36827\%, respectively. Black curve reports the fit $\lambda_{min}\propto\sqrt{\gamma_c-\gamma}$. The energy drops are reported in blue dots.}
\label{fig:eigv}
\end{figure}

In Fig.~\ref{fig:displacement_snap}a the energy vs strain curve is reported. There we identify two events -- marked by circles and magnified in the insets -- which are of different nature, one being associated to a typical quasi-localized rearrangement without any change in the atomic coordination, the other resulting from the first observed bond breaking.
Thanks to these rearrangements, some of the internal stress is released, and a consequent drop in the energy occurs. As shown in the insets using the smaller $\delta\gamma$ values, both events manifest a drop in the total energy, which results of about 0.9\,eV and 5.4\,eV for rearrangement and bond breaking respectively, in line with previous works~\cite{koziatek2015short}.
The non affine atomic displacements corresponding to the energy drops are represented in Fig.~\ref{fig:displacement_snap}b and \ref{fig:displacement_snap}c (the corresponding movies are available in Supplemental Material \cite{Note21}). We note that while the rearrangement event consists in displacements along multiple directions, the bond breaking produces displacements mainly along the principal strain direction $z$. Specifically, the rearrangement involves changement in angle in two under-coordinated silicon atoms, and the bond breaking occurs between a Si and a O$^{3-}$ atom. Therefore both events appear in the presence of a structural defect.

The number of particles involved in such fundamental non-affine events can be estimated by the participation number $P_n=\sum_{i=1}^N{(u_i/u_{max})^2}$, with $u_i$ the displacement modulus of atom $i$, and $u_{max}$ the maximum $u_i$. Such calculation for rearrangement and bond breaking events gives $P_n^{RR}=3.73$ and $P_n^{BB}=6.11$, respectively, revealing that the more energetic event involves, as expected, a larger effective number of particles.
Furthermore, the participation ratio $P_r=\sum_{i=1}^N{(\B e_i\cdot \B e_i)^2}/[\sum_{i=1}^N{(\B e_i\cdot \B e_i)}]^2$, calculated using the Hessian eigenvectors $\B e_i$ right before the critical strain, results $P_r^{RR}=0.31$ and $P_r^{BB}=0.20$.

Analytical investigation of the rearrangement events induced by external stress can be performed by computing the $\B H$ matrix, and by following the direction of the softest mode.
The results of this investigation are reported in Fig.~\ref{fig:eigv}, showing that for both selected events the smallest eigenvalue $\lambda_{min}$ progressively decreases following a square-root trend, and vanishes at the critical strain value $\gamma_c$ marking a saddle point in the energy configuration space. 
As in the case of metallic glasses, governed by isotropic interactions, in which stress release is associated to a irreversible plastic event, we have verified that the same irreversibility occurs in the covalently bonded system under consideration. The application of a negative strain rate after a rearrangement does not follow the configurational path that led the system to that rearrangement.

\begin{figure}[t]
\centering
\includegraphics[width=\columnwidth]{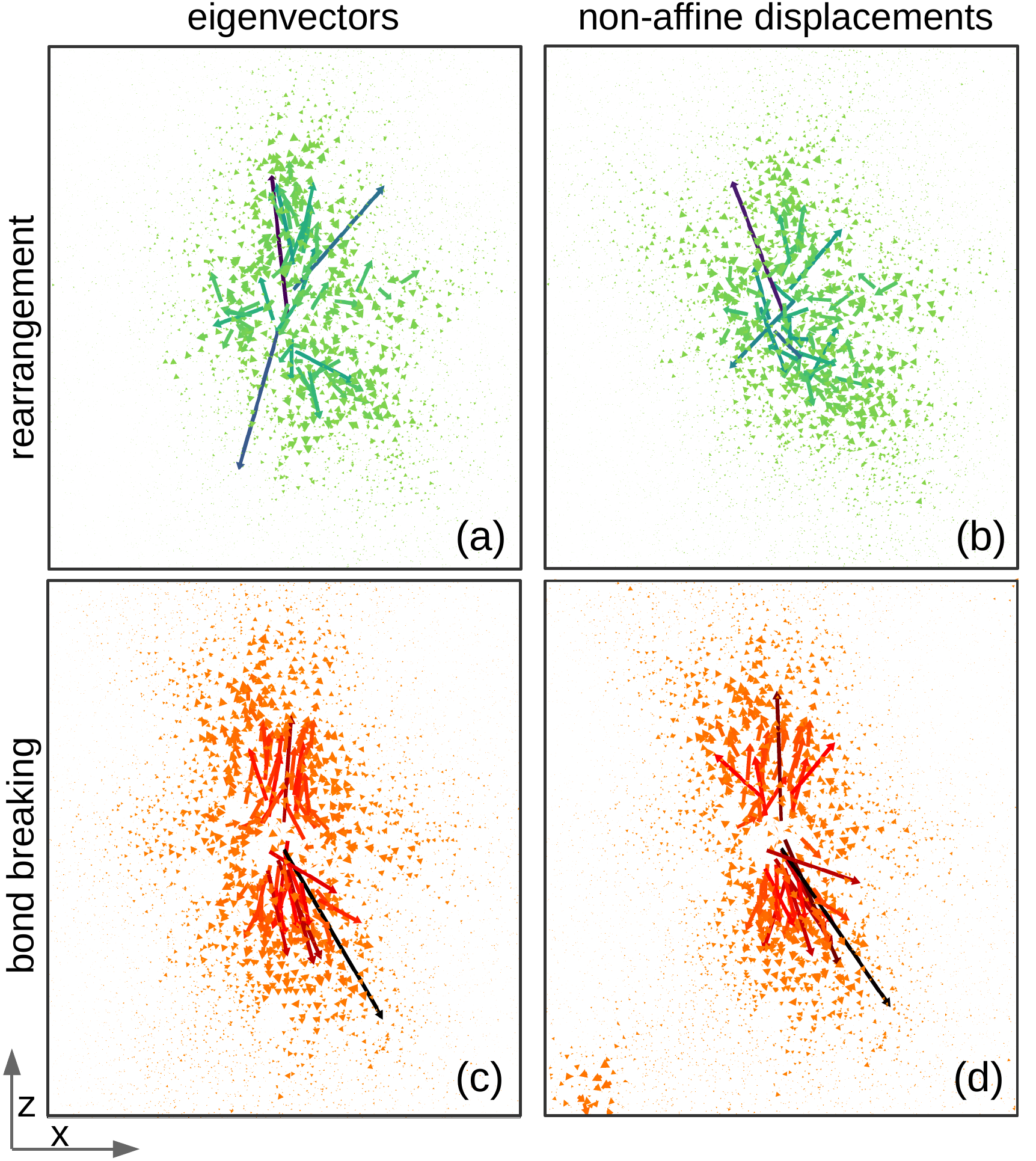}
\caption{Comparison between normalized non-affine displacements (left) and eigenvectors (right) of the configuration before the bond breaking or energy drop: (AQS$_2$ steps). Arrows are colored with respect to the modulus of the vectors. Arrows have been rescaled by a factor of 50, the view is ``orthogonal''. Panels (b) and (d) are related to Figure~\ref{fig:displacement_snap}(b) and (c), right panels.}
\label{fig:eigvec}
\end{figure}

\begin{figure}[t]
	\includegraphics[width=\columnwidth]{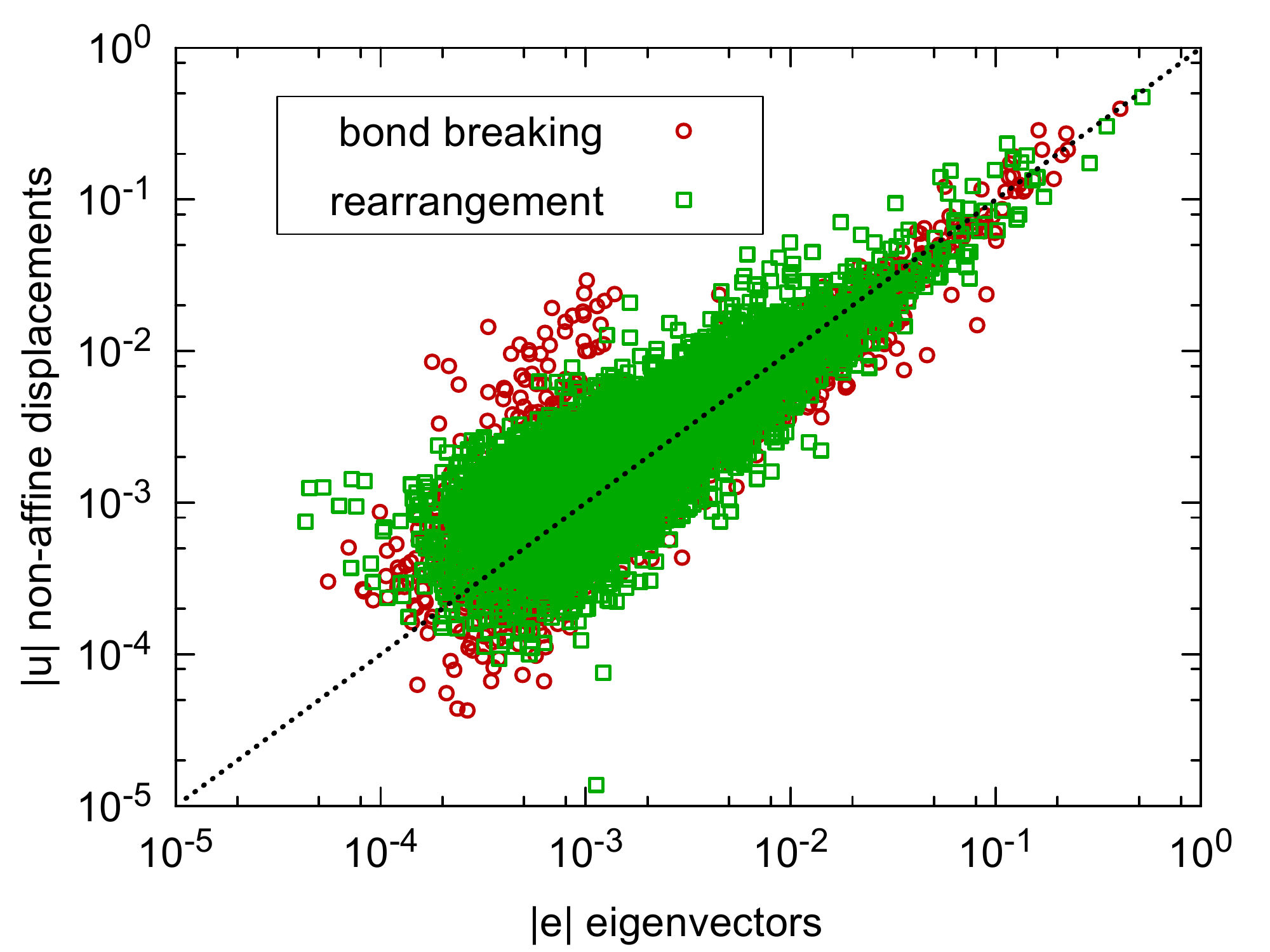}
	\caption{Eigenvectors vs non-affine displacement comparison for the rearrangement (squares) and bond breaking (circles) events, compared to the ideal correlation line (dotted line).}
	\label{fig:fig_4}
\end{figure}

Further information can be obtained by analyzing the eigenvectors associated with the lowest eigenvalue $\lambda_{min}$, to be compared with the non-affine displacement fields, calculated between the frame after the energy drop and the one before. \\
In Figure~\ref{fig:eigvec} we present such comparison for the selected plastic events showing an almost exact matching, especially in the case of bond breaking: The scalar product, $s=\sum_{i=1}^{N}{(\B u_i \cdot \B e_i)}$, of the normalized  eigenvectors $\B e_i$ and non affine displacements $\B u_i$ results $s_{RR}\simeq 0.74$ and $s_{BB}\simeq 0.91$ respectively.
The higher correlation in the latter case is likely due to the fact that bond breaking occurs along the strain direction, while the rearrangement occurs through a local rotation of bonds, i.e.\ not connected to the strain direction.
In any case, both mechanical events show a high correlation between the eigenvectors and displacements. This evidence is also supported by Figure~\ref{fig:fig_4} in which the individual $e_i$ vs $u_i$ moduli are compared.

In summary, we have analyzed, using Hessian methods, the nature of non-affine responses to mechanical shear strain in a model of silica.
The difference from other examples of similar studies is that in silica we have {\em directional chemical  bonds} between atoms, and these can be broken.
In most models of glass formers one cannot assign actual bonds, and one can even discuss glass physics with repulsive interaction only. The presence
of bonds enriches that discussion of non-affine responses to strain, offering plastic events that do not exist in most of the studied models
of glass formers. We could therefore identify two distinctly different non-affine responses in silica, one that corresponds to other models with
so called T1 processes that involves particles moving out and particles moving in on a quadrupolar quasi-localized structure, but also an elementary
event of bond breaking. Both events are accompanied by an eigenvalue of the Hessian approaching zero with a square-root singularity and are
associated with a stress and energy drop. Even in the case of bond breaking the system response is again quadrupolar. This result is important for theoretical modeling of
failure in real amorphous materials since our understanding of amorphous plasticity has often
relied on mesoscale models assuming that plastic deformation can be decomposed into
a series of discrete localized plastic instabilities \cite{nicolas2018deformation}. While
this assumption was supported by atomistic simulations in simplified isotropic models
for glasses \cite{12DKP}, the present study shows that the same description holds for more
realistic anisotropic models. Finally, an important aspect of our findings is that using the eigenfunction associated with the lowest eigenvalue one can predict the locus of the non-affine
response \cite{10KLLP} even in realistic anisotropic conditions as those simulate here.

\acknowledgements{This work is supported by the cooperation project COMPAMP/DISORDER jointly
funded by the Ministry of Foreign Affairs and International Cooperation (MAECI) of Italy
and by the Ministery of Science and Technology (MOST) of Israel.}

\bibliographystyle{unsrt}
\bibliography{biblio}

\end{document}